# Classification of signaling proteins based on molecular star graph descriptors using Machine Learning models

Carlos Fernandez-Lozano[a,*], Rubén F. Cuiñas[a], José A. Seoane[b], Enrique Fernández-Blanco[a], Julian Dorado[a], Cristian R. Munteanu[a,c]

[a] Information and Communications Technologies Department. Faculty of Computer Science, University of A Coruna, Campus de Elviña s/n, 15071 A Coruña, Spain

[b] Bristol Genetic Epidemiology Laboratories, School of Social and Community Medicine, University of Bristol, Oakfield House, Oakfield Grove, Bristol BS82BN, UK

[c] Department of Bioinformatics - BiGCaT, Maastricht University, P.O. Box 616, UNS50 Box 19, NL-6200 MD, Maastricht, The Netherlands

carlos.fernandez@udc.es, ruben.fcuinas@udc.es, j.seoane@bristol.ac.uk, efernandez@udc.es, julian@udc.es, crm.publish@gmail.com

**Abstract**

Signaling proteins are an important topic in drug development due to the increased importance of finding fast, accurate and cheap methods to evaluate new molecular targets involved in specific diseases. The complexity of the protein structure hinders the direct association of the signaling activity with the molecular structure. Therefore, the proposed solution involves the use of protein star graphs for the peptide sequence information encoding into specific topological indices calculated with S2SNet tool. The Quantitative Structure – Activity Relationship classification model obtained with Machine Learning techniques is able to predict new signaling peptides. The best classification model is the first signaling prediction model, which is based on eleven descriptors and it was obtained using the Support Vector Machines - Recursive Feature Elimination (SVM-RFE) technique with the Laplacian kernel (RFE-LAP) and an AUROC of 0.961. Testing a set of 3114 proteins of unknown function from the PDB database assessed the prediction performance of the model. Important signaling pathways are presented for three UniprotIDs (34 PDBs) with a signaling prediction greater than 98.0%.

**Keywords**

Feature Selection; SVM-RFE; Topological Indices; Signal Transduction Pathway

*Corresponding author: Dept. of Information and Communication Technologies, Computer Science Faculty, University of A Coruña 15071, Spain; Email: carlos.fernandez@udc.es; Tel: (+34) 981167000, Ext. 1302, Fax: (+34)981167160.

**Introduction**

Responding to immediate changes in the environment is essential for the survival of cells. A cell is able to receive many signals simultaneously and to integrate the information into a unified action plan. In addition, a cell can also send signals to the environment. These signals may be chemical, mechanical, thermal or optical (Rhee, 2006).

Examples of chemical signals are growth factors, hormones, neurotransmitters, and extracellular matrix components. Due to the fact that any molecule can travel at distance, these signals may have a local or long distance effect (Jordan et al., 2000). An example of short-range signaling molecules is that of the neurotransmitters. They can travel a tiny distance between adjacent neurons or between neurons and muscle cells. On the other hand, an example of long-range molecules is that of the follicle-stimulating hormone, which travels from the mammalian brain to the ovary, in order to trigger the egg release. Sensory cells in the skin, ear, and human vascular system could receive mechanical stimuli.

A cell can receive and modulate the signals through hundreds of protein receptor types (Wong et al., 2012). They cause a physiological response specific for different types of molecules. The receptors for dopamine are different than those for insulin, and others can even respond directly to light or pressure. Receptors are generally transmembrane proteins, with an outside part which binds to specific signaling molecules and with an internal part that can initiate specific signaling pathways in the cell.

The membrane receptors could be grouped into three major classes: G-protein-coupled receptors (Kobilka, 1992; Rohrer and Kobilka, 1998), ion channel receptors, and enzyme-linked receptors (Evans and Levitan, 1986). Therefore, these receptors transform external signals into internal ones via protein action, ion channel opening, or enzyme activation, respectively. Due to the internal and external part of the receptors, they allow cellular actions of an external molecule without entering the cell of the signal molecules. Other receptors are localized deep inside the cell, even in the nucleus and they may bind to molecules which are able to pass through the plasma membrane (ex: gases, steroid hormones) or that are products of the internal metabolism (Dykstra et al., 2003).

The receptors can launch a series of biochemical reactions within the cell due to the conformational changes determined by the signal interaction. These signal transduction cascades can amplify the message, producing multiple intracellular signals. In the first step, the activation of the receptors can trigger the synthesis of small molecules called second messengers, such as cyclic AMP (cAMP), which initiate and coordinate intracellular signaling pathways (Sassone-Corsi, 2012). The activation of adenylyl cyclase produces the synthesis of hundreds/thousands of cAMPs that activate the enzyme protein kinase A (PKA), which then phosphorylates multiple protein substrates. The cAMP signaling stops when cAMP is degraded by the enzyme phosphodiesterase.

The signaling proteins are the best targets for drugs in order to modify any biological activity. Therefore, searching for new proteins with signaling function is essential. Due to the fact that the experimental methods are expensive and time-consuming, the theoretical methods could offer the practical solution for this screening. Therefore, classification models that link the protein structure to the signaling activity could be obtained using Machine Learning techniques. The molecular information can be encoded into invariant molecular descriptors based on molecular graph topology, 3D protein conformation, peptide sequence and physical-chemical properties of the amino acids.

The classification model represents a Quantitative-Structure-Activity-Relationship (QSAR) (Archer, 1978) between the protein structure and the biological function. The QSAR models (Puzyn et al., 2010) were extensively used for antifungal (Gonzalez-Diaz et al., 2006), antiviral (Prado-Prado et al., 2011b), and antimalarial (Katritzky et al., 2006) drugs and they were extended to macromolecules such as proteins (González-Díaz and Uriarte, 2005; González-Díaz et al., 2009; Ivanciuc, 2009; Randic et al., 2007; Randić et al., 2006) or nucleic acids (Gonzalez-Diaz et al., 2005; Li and Wang, 2004; Randić et al., 2000). Previous studies on other protein functions were focused on anti-oxidant (Fernandez-Blanco et al., 2012), transporter (Fernandez-Lozano et al., 2013), enzyme regulator (Concu et al., 2009; Fernandez-Lozano et al., 2014b), cell death-related (Fernandez-Lozano et al., 2014a) or cancer-related (Aguiar-Pulido et al., 2012; Munteanu et al., 2009) proteins. There are also other works which involves QSAR models for the prediction of protein inhibitors, where the structures or sequence of the proteins have been considered (Prado-Prado et al., 2012; Prado-Prado et al., 2011a; Speck-Planche et al., 2012; Speck-Planche et al., 2013; Viña et al., 2009)

The scientific community pays special attention to signaling protein classification problems and further efforts are needed in order to complete the scientific knowledge in this field, as the classification of proteins is a key step in computational genomics. Simple Machine Learning approaches were used in the past to automate this process: for protein classification (Ahmad et al., 2014), classification of proteins known to be frequently mutated in human cancer (U et al., 2014), classification of proteins in the early clinical stage of rheumatoid arthritis (Pratt et al., 2012) or to explore the non-linear relationships between mitochondrial morphology and apoptotic signaling (Reis et al., 2012). Those Supervised Machine Learning techniques for signaling protein classification are mainly Artificial Neural Networks, Support Vector Machines, Naïve Bayes, Decision Trees or Random Forest. Even though the use of Machine Learning approaches for protein classification is a common topic, it is difficult to compare the performance of newer approaches. This is due mainly to the lack of information about the sequences, date of extraction, concrete database or information about the configuration of the techniques in publications. In order to get a reproducible experimentation all this information was in the current paper and the datasets were made available at http://dx.doi.org/10.6084/m9.figshare.1330132.

The aim of the current study is to obtain a classification model for signaling protein prediction based on star graph descriptors of protein sequences and Machine Learning techniques. The current paper is organized as follows: the Materials and Methods section describes the methodology and the particular methods used in this work, as well as the process to generate our dataset; the Results and Discussion section includes a comparison of the proposed algorithm with the above-mentioned state-of-the-art algorithms for protein classification and an experimental analysis of the performance of the proposed method is performed, as this is a crucial and necessary task in any research study (García et al., 2010); finally, the conclusions are presented.

**Materials and methods**

The flowchart of the current study methodology is presented in **Figure 1**. Two classes of sequences are evaluated: signaling and non-signaling protein groups. The database consists of amino acid sequences (primary structure) of signaling and non-signaling proteins in FASTA format. In the second step, the sequences of amino acids are transformed with S2SNet (Munteanu et al., 2009) into topological indices of the protein Star graphs. The descriptors of the graphical representation of the protein sequence are then used to search for the best QSAR classification model that will predict the signaling molecular function for new amino acid sequences. Machine learning techniques are widely used with biological problems in order to predict protein complex biological functions as mentioned before or more recently for protein-protein interactions (Munteanu et al., 2015).

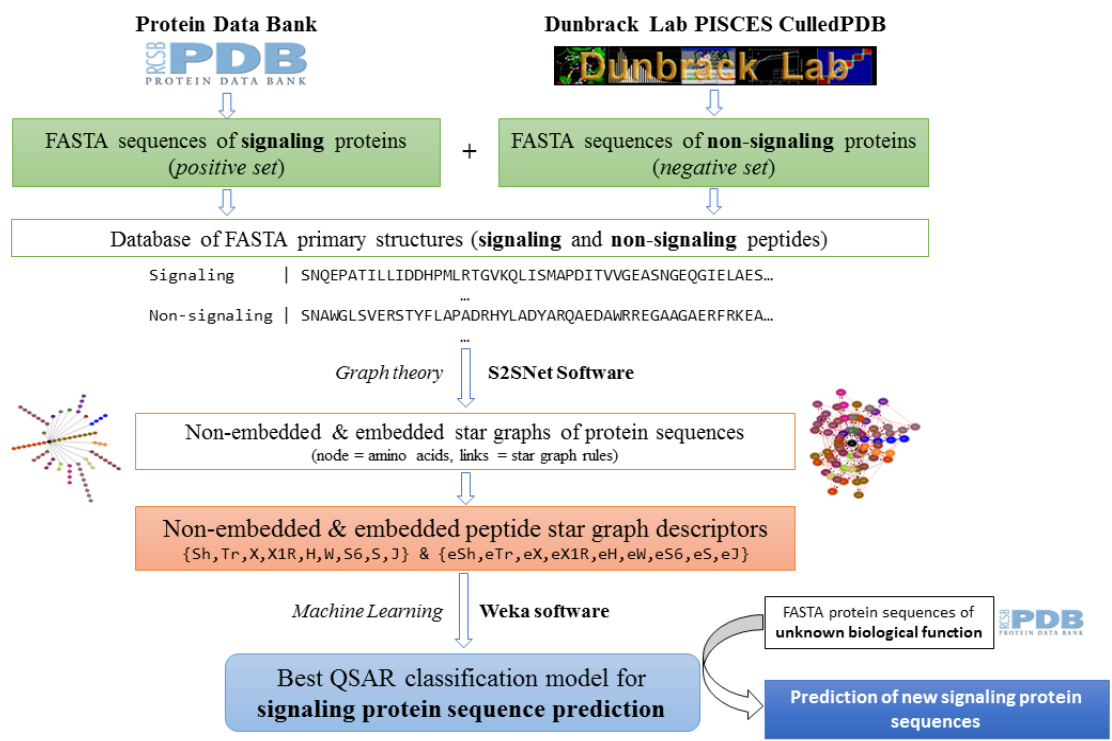

**Figure 1.** Methodology flowchart including molecular graphs and Machine Learning for signaling protein classification model

The authors have tested the different classification techniques using 10-fold cross-validation to split the data (McLachlan et al., 2004). Therefore, the dataset is randomly partitioned into 10 equal-sized bins. 9 bins were picked 10 times to train the models and the remaining bin was used to test them, each time leaving out a different bin.

The performance of prediction models for the signaling and non-signaling classes is evaluated using a confusion matrix. The following accuracy measures are typically used for two-class classifiers (Ferri et al., 2009): classification rate, precision, sensitivity, specificity, F-measure, and Area Under the Receiver Operating Characteristic Curve (AUROC). The higher the precision, the less effort is required in testing and inspection; and the higher the recall, the fewer defective modules go undetected. However, there is a trade-off between precision and recall and, therefore, a combination of both is needed in a single efficiency measure, known as F-measure, which considers both precision and recall equally important (Witten and Frank, 2005). AUROC is considered a better measure than accuracy when comparing classifiers (Jin, 2005).

We started our experiments using the Weka software (Hall et al., 2009) in order to establish the baseline AUROC performance and be able to compare our experiments with complex feature selection approaches using the R package (Development Core, 2005) and our particular implementation of a Particle Swam Optimization algorithm. The aim of these techniques is to find the best minimum combination of features with the highest performance in order to avoid noisy and irrelevant features. There are mainly three different feature selection approaches: filter, wrapper and embedded (Fernandez-Lozano et al., 2015; Saeys et al., 2007). In this work we tested the well-known wrapper (treat the problem in the same step as the model selection) and embedded (inside the classification model) approaches. Several methods will be used to select the minimum number of features for the best QSAR model.

Finally, the proposed model for signaling protein classification is available as a free Web tool at http://bio-aims.udc.es/Signal-Pred.php in order to be validated by the scientific community and we used it in order to test the capability of the model to predict signaling proteins from a list of 3114 proteins of "unknown" function from the PDB Databank, results are in the Results and Discussion section of this paper.

*Protein sets*

The positive group of 608 signaling protein sequences was downloaded in FASTA format from the Protein Databank (Berman et al., 2000) using the "Molecular Function Browser" in the "Advanced Search Interface", "Signaling (GO: 0023052)", protein identity cut-off = 30%. The negative group of 2077 non-signaling proteins was downloaded from the PISCES CulledPDB (http://dunbrack.fccc.edu/PISCES.php) (Wang and R. L. Dunbrack, 2003) (November 19$^{th}$, 2012) using sequence identity (degree of correspondence between two sequences) less than 20%, resolution of 1.6 Å and R-factor 0.25. This negative set is representing all the other protein structures from the Protein Databank. Before any data processing, the negative set was verified to not contain any protein chain from the positive set (any signaling protein). Thus, two sets of protein chain sequences have been obtained: the signaling sequences and the representative proteins without signaling molecular function (in Protein Databank).

This kind of unbalanced data is not the most suitable to be used as an input for learning algorithms because the results present a high sensitivity and low specificity; learning algorithms tend to classify most samples as part of the most common group (Fernández-Navarro et al., 2011). To avoid this situation, a pre-processing stage is needed in order to get a more balanced dataset, in this case by means of the synthetic minority oversampling technique (SMOTE) (Chawla et al., 2002). In short, SMOTE provides a more balanced dataset using an expansion of the lower class by creating new samples, interpolating other minority-class samples. After this pre-processing, the final dataset is composed of 1824 positive samples (signaling protein chains) and 2432 negative cases (non-signaling protein chains).

These protein sequences have been processed with the S2SNet application in order to obtain the different topological indexes used in this study. From each sequence, 42 attributes are calculated for the corresponding embedded/non-embedded Star Graph as shown in **Table 1**. The variables are grouped in accordance with the embedded and non-embedded Star graphs. Features have also been grouped according to different subsets: one called *Sh*, which includes the attributes related to entropy; a subset called Tr, which includes attributes related to the traces; a subset called *X*, which includes the attributes related to the polygon indexes and finally the *remaining* attributes regarding the general shape of the graphs. Several methods from Weka and R have been used to find the best signaling classification methods.

In the next step, the protein sequence descriptors are calculated using molecular graphs. These numbers will be used to search for the best QSAR model to predict signaling protein sequences.

**Table 1.** Embedded/non-embedded attributes

|  | Non-embedded | Embedded |
|---|---|---|
| Sh | *Sh0*, *Sh1*, *Sh2*, *Sh3*, *Sh4*, *Sh5* | eSh0, eSh1, eSh2, eSh3, eSh4, eSh5 |
| Tr | Tr0, Tr2, Tr4 | eTr0, eTr2, eTr3, eTr4, eTr5 |
| X | X0, X1R, X2, X3, X4, X5 | eX0, eX1R, eX2, eX3, eX4, eX5 |
| Remaining | H, W, S6, S, J | eH, eW, eS6, eS, eJ |

*Star Graph Topological Indices*

The protein primary sequences have been transformed into Star Graphs using amino acids as vertices (nodes) connected by star graph rules or by the initial connectivity into the protein primary sequence (peptide bonds). Star Graph is a special type of tree with *N* vertices, where one vertices has *N*-1 degrees of freedom and the remaining *N*-1 vertices have a single degree of freedom (Harary, 1969).

This type of graph is built by adding all the amino acids into 23 possible branches ("rays"), all types of amino acids in one specific branch (according to their order in the sequence). The star centre is a non-amino acid vertex (Randic et al., 2007). 20 groups have been used for the standard amino acids (A, R, N, D, C, E, Q, G, H, I, L, K, M, F, P, S, T, W, Y, V), 2 groups with two nonstandard ones (U, O) and 1 group with four ambiguous amino acids (B, Z, J, X). Therefore, the following information of the protein sequence is encoded into star graph descriptors: amino acid type, order in the sequence, and frequency in branches.

Different forms of graphs, associated with distinct distance matrices, can represent a protein. The standard method to create a Star Graph consists of the next steps: each amino acid/vertex holds the position in the original sequence and the branches are labeled alphabetically by the 3-letter amino acid code. The embedded graph contains, in addition to the connectivity from the non-embedded graph, the initial protein sequence connectivity.

The star graph descriptors are calculated using the graph connectivity matrix, the power of the matrices (Markov chain transitions), distance matrices into a graph (length of the pathways between all nodes in the star graph), and node degrees (amino acid connections in the graph). The connectivity matrix is normalized and represents the basis of these descriptors. The used tool was Sequence to the Star Networks (S2SNet) (Fernandez-Lozano et al., 2013), an application designed using Python/wxPython (Rappin and Dunn, 2006) and *Graphviz* (Koutsofios and North, 1993) as a plotting back-end. TIs are calculated using the embedded and non-embedded Star graphs, without weights, using Markov normalization and a power of matrices/indices (*n*) up to 5.

The complete formulas for all the S2SNet TIs (Todeschini and Consonni, 2002) used in the current study are given in the references (Fernandez-Blanco et al., 2012; Fernandez-Lozano et al., 2013) presented: Shannon entropies ($Sh_n$), trace of connectivity matrices ($Tr_n$) with n = 0-5, Harary number (H), Wiener index (W), Gutman topological index (S6), Schultz topological index (nontrivial part) (S), Balaban distance connectivity index (J), Kier-Hall connectivity indices (Xn) with n=0, 2-5, Randic connectivity index (X1R). The embedded descriptors have the prefix "e".

*Machine learning classification methods*

The authors have performed several experiments in order to select the best model. The classification implementations used in those test were the ones included in the well-known Machine Learning library Weka (Hall et al., 2009). More specifically, the authors have used: Naïve Bayes (NB) (John and Langley, 1995), Random Forest (RF) (Breiman, 2001), J48 (Witten and Frank, 2005) (the Weka implementation of c4.5 algorithm), LibLINEAR (Fan et al., 2008), SVM (Chang and Lin, 2011) and K* (Cleary and Trigg, 1995). Among all these classifiers, SVM was chosen because of the reasons presented in the next section of the paper.

Supervised learning algorithms for classification problems such as Support Vector Machines (Vapnik, 1979; Vapnik, 1995) (SVM) or Random Forest (Breiman, 2001) (RF) are commonly used for complex protein function prediction. For binary classification problems, SVMs are designed to find the largest margin separation hyperplane between two classes, as shown in **Figure 2**. The position and orientation of this hyperplane is defined by the support vectors (Burges, 1998). The hyperplane can be defined as $f(x) = \langle w, x \rangle + b = 0$ where $x$ represents $n$ input vector, the output vector is $y \in \{-1, 1\}$, $b$ is the bias and $w$ is the weight vector.

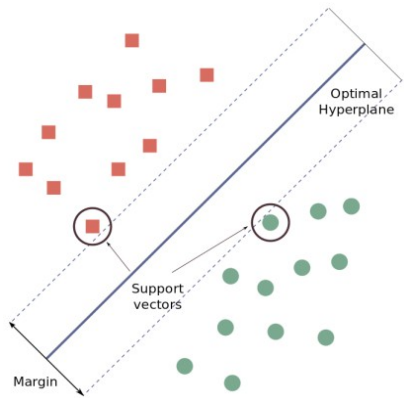

**Figure 2.** The large-margin hyperplane separates the two classes (support vectors in black circles)

Real problems are usually not linearly separable and SVMs handle these problems by an implicit mapping to a new and higher dimensional space by means of a kernel function, where data become linearly separable. A kernel function $K$ calculates the inner product of two vectors $x, x'$ in a given feature mapping $\Phi : X \to H$. There are several different types of linear and nonlinear kernel functions such as: linear (1), Gaussian radial basis (RBF) (2) or Laplace radial basis (LAP) (3).

$$K(x, x') = \langle x, x' \rangle \quad (1)$$

$$K(x, x') = \exp(-\sigma \|x - x'\|^2) \quad (2)$$

$$K(x, x') = \exp(-\sigma \|x - x'\|) \quad (3)$$

Random Forest algorithm is based on the ensemble of a particular number of unpruned decision trees with randomly selected features. Each tree gives a vote according with classification performance and the forest chooses the most voted trees.

*Feature selection approaches*

Feature selection (FS) techniques emerge to deal with high-dimensional feature spaces in order to remove noisy and redundant features to improve classifier performance. The number and relevance of input variables can affect the classifier performance. The idea behind this technique is to find the best subset of input features. This subset should describe the structure of the data better than or equal to the original set of features. There are mainly three major categories of FS techniques (Saeys et al., 2007): wrapper, filter and embedded.

Particle Swarm Optimization (Kennedy and Eberhart, 1995) (PSO) is a bio-inspired optimization algorithm based on the simulation of the social behavior of bird flocks. In the last fifteen years PSO has been applied to a very large variety of problems (Poli, 2008) and numerous variants of the algorithm have been presented (Banks et al., 2007). It was initially developed for continuous problem solving but later on, it was modified for discrete problems. This approach was known as binary PSO.

During the execution of PSO, a set of particles moves within the function domain searching for the optimum of the function (best fitness value). The motion of each particle is driven by the best positions visited so far by the particle itself (pbest) and by the entire swarm (gbest) or by some pre-defined neighborhood of the particle (lbest). Consequently, each particle relies both on "individual" and on "swarm" intelligence, and its motion can be described by two simple discrete-time equations, which regulate the particle's position (4) and velocity (5):

$$\vec{X}_i^{t+1} = \vec{X}_i^t + \vec{V}_i^t \quad (4)$$

$$\vec{V}_i^{t+1} = \omega \vec{V}_i^t + c_1 \vec{R}_1 \left( pbest - \vec{X}_i^t \right) + c_2 \vec{R}_2 (lbest - \vec{X}_i^t) \quad (5)$$

Each particle (i) position is updated every iteration (t) in the direction of its velocity. The velocity itself varies towards the direction of pbest and lbest, affected by the cognitive $(c_1)$ and social $(c_2)$ factors, by a uniformly distributed vector $\vec{R}$ and by the inertia weight $(\omega)$, as shown in **Figure 3**.

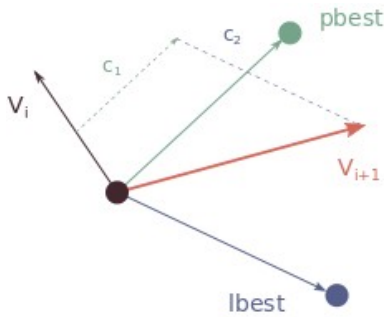

**Figure 3.** Velocity vector update

PSO could be used in combination with a classifier for FS following a wrapper approach. Based on the latest Standard PSO implementation (SPSO-2011) (Central, 2014; Clerc, 2012), we modified it in order to add a binary representation for each particle as a feature mask for the input feature space and a real-number representation of the parameters of the kernel. SVM decision function is obtained by LIBSVM (Chang and Lin, 2011). Furthermore, we are able to perform a feature selection process while we are searching within the same process for the best combination of parameters. Thus, different criteria are optimized simultaneously in a multi-objective approach.

Support Vector Machines Recursive Feature Elimination (SVM-RFE) is one of the most successful classification algorithms for feature selection following an embedded approach. It was initially introduced using SVMs (Guyon et al., 2002) with linear kernel, but other classifiers could be used.

In this work, the R Statistical Package (Development Core, 2005) has been used to conduct RFE with several different functions, among other packages (more specifically the Caret (Kuhn, 2008), Kernlab (Karatzoglou et al., 2004) and pROC (Robin et al., 2011) packages for graphic purposes). The Caret package (the acronym for Classification and Regression Training) provides a set of functions to perform a backward selection of features based on score ranking. Caret package capacities had to be enhanced in order to include a new ranking criterion for supporting SVM-RFE as initially proposed by Guyon (Guyon et al., 2002).

Our improved version of SVM-RFE has been modified using $\|w\|^2$ as ranking criterion for measuring the importance of a particular feature instead of the $w$ proposed by Guyon for linear kernels. In the case of RFE-RF, the importance of each feature is calculated based on the RF (Breiman, 2001) proposed by Breiman. The classification accuracy is calculated at the beginning of the process. A feature is removed completely from the RF, not only in a particular tree, and accuracy is checked again. The importance of this feature is calculated based on the accuracy difference.

**Results and discussion**

The final dataset is composed of 1824 positive samples (signaling protein chains) and 2432 negative cases (non-signaling protein chains). These protein sequences have been processed with the S2SNet application in order to obtain the 42 different topological indexes used in this study. Six different classification models have been used in order to generate the baseline performance. After this, four different complex approaches for feature selection have been tested in order to improve baseline best model's results.

After an experimental analysis comparison study, we concluded that RFE-LAP is, according with a null hypothesis test, statistically better than the others.

*Baseline techniques without feature selection*

The experiments are started using six state-of-the-art Machine Learning techniques for protein classification in order to establish the baseline performance. However, further experimental analysis of the performance of the proposed method should be calculated. SVM[a] with RBF kernel yields the best results with AUROC values of 0.948. Comparing these results with the rest of the tested techniques, it was found that it improved significantly the results offered by LibLINEAR[f] (0.657), NB[e] (0.619) and J48[d] (0.687). Moreover, it also improved the results of RF[c] (0.935) and K*[b] (0.939), but at a smaller extent. ROC curves for reference models are shown in **Figure 4**.

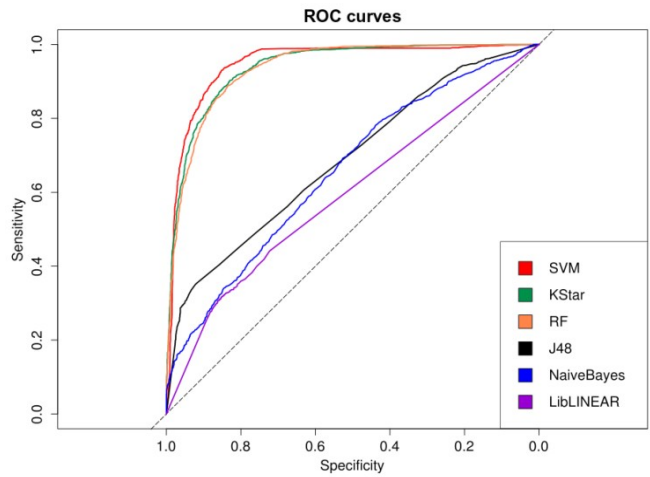

**Figure 4.** ROC curves for reference models

[a] weka.classifiers.functions.LibSVM -S 0 -K 2 -D 3 -G 32.0 -R 0.0 -N 0.5 -M 100.0 -C 16.0 -E 0.001 -P 0.1 -B -seed 1

[b] weka.classifiers.lazy.KStar –B 20 –M a

[c] weka.classifiers.trees.RandomForest –l 500 –K 5 –S 1 –num-slots 1

[d] weka.classifiers.trees.J48 –C 0.25 –M 2

[e] weka.classifiers.bayes.NaiveBayes –K

[a] weka.classifiers.functions.LibLINEAR –S 7 –C 10.0 –E 0.01 –B 1.0 -P

*Machine learning techniques with Feature selection*

In Bioinformatics, it is usual to perform a feature selection process in order to reduce as much as possible the number of features and to improve results of the algorithms. Four different classification models for Feature Selection have been tested: a Particle Swarm Optimization with an SVM as decision function with RBF kernel (PSO-SVM), RFE-Support Vector Machines with RBF (RFE-RBF) and Laplacian kernels (RFE-LAP) and RFE-Random Forests (RFE-RF). In order to validate the models the experiments have been repeated 5 times, as shown in the error bar plot of **Figure 5**.

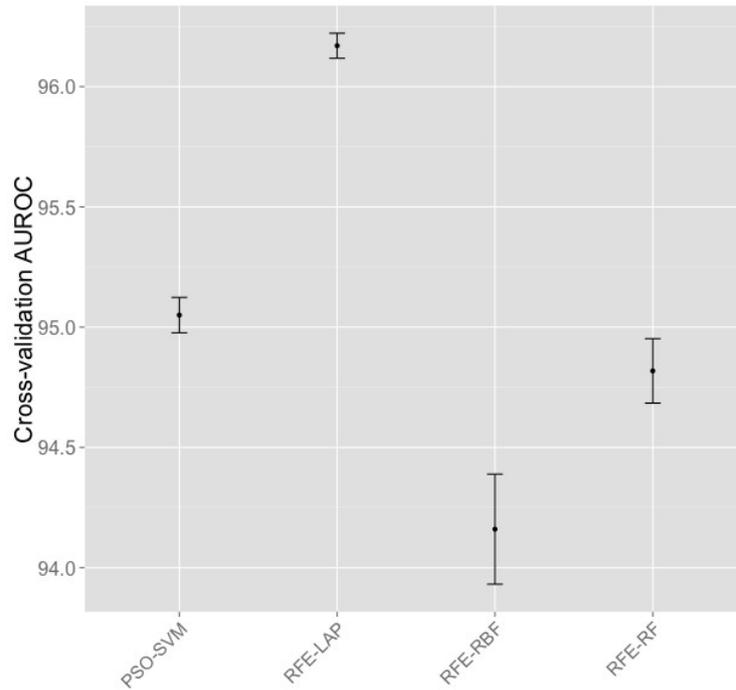

**Figure 5.** Results achieved for feature selection techniques

The AUROC with no feature selection has been previously estimated using SVM and a value of 0.948 was obtained. These results were improved with recursive feature selection (achieving 0.961 ± 0.11%) using RFE-LAP (11 features: X5, H, S6, X4, eS6, eH, eTr3, eX5, eTr5, eTr0, Tr0) and (0.948 ± 0.29%) using RFE-RF (7 features: eTr3, eTr5, J, eTr4, eX5, X5, eTr2) and with a bio-inspired hybrid approach (achieving 0.950 ± 0.16%) using PSO-SVM (30 features: Sh0, Sh1, Sh2, Sh3, Sh4, Tr2, H, W, J, X1R, X4, X5, eSh0, eSh1, eSh2, eSh3, eSh4, eSh5, eTr0, eTr2, eTr3, eTr4, eTr5, eW, eS, eJ, eX0, eX1R, eX4, eX5).

*Experimental analysis and best model determination*

Despite the fact that ROC curves are a powerful tool for visualizing the performance of a classifier, they are not commonly accepted as a means to compare different techniques performing the same classification task (Hanczar et al., 2010; Hand, 2009; Lobo et al., 2008). To find out whether these results are statistically significant, a null hypothesis test is performed next for the analysis of the behavior of those techniques (García et al., 2009; García et al., 2010; Sheskin, 2011). Unfortunately, parametric tests are commonly used to compare different models without checking if the required conditions are fulfilled (independence, normality and heteroscedasticity) when applying those tests. The independence condition is fulfilled when the occurrence of an event does not modify the probability of another event. A Gaussian distribution (mean and variance with a particular value) is assumed with the normal assumption. Finally, the heteroscedasticity indicates the existence of a violation of the hypothesis of equality of variances.

In our experiments, the data were separated using a 10-fold cross validation approach. Thus, the independence condition is fulfilled because of the independent nature of the samples. A normality analysis was performed using the Shaphiro-Wilk test (Shapiro and Wilk, 1965) with a level of confidence $\alpha = 0.05$. The null hypothesis was not rejected with values $W=0.9404$ and $p-value<0.244$ therefore, considerate can be stated that these results follow a normal distribution. **Figure 6** shows the quantile-quantile plot.

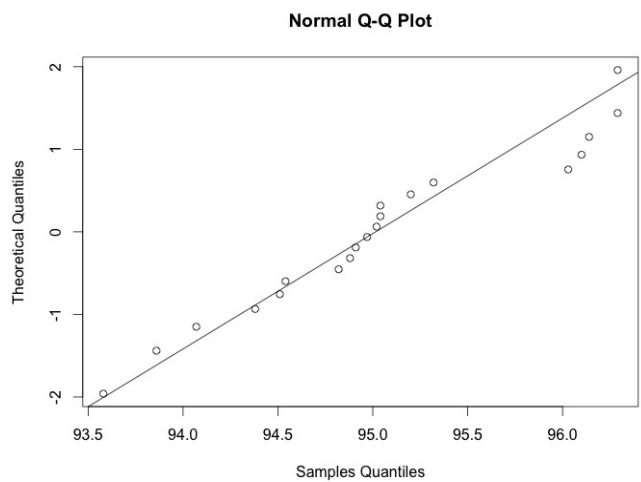

**Figure 6.** Q-Q Plot of observed versus expected values

In order to evaluate heteroscedasticity a Barlett test (Bartlett, 1937) was performed, with a level of confidence $\alpha=0.05$. The null hypothesis was rejected with values Bartlett's K-Squared = 8.5804 and $p-value<0.03542$ thus it can be stated that our results are heteroscedastic. One of the three conditions for using a parametric is not fulfilled, thus a non-parametric test, such as Friedman (Friedman, 1937), Friedman Aligned Ranks (Hodges and Lehmann, 1962) or Quade (Quade, 1979), should be performed in order to rank these models. These tests can be used under the same circumstances but in (García et al., 2010) authors stated that Quade test should be used when the number of algorithms to compare was low. This test performs a ranking of the models compared, assuming as null hypothesis that all models have the same performance. The average rankings of the techniques using a Quade test are shown in **Table 2** with the following values: Quade statistic = 23.91 and $p-value<2.37\times 10^{-5}$ distributed with 3 and 12 degrees of freedom $F(3,12)$, thus the null hypothesis is rejected with a high level of significance, showing that the best model is RFE-LAP.

**Table 2.** Quade's average ranking

| Technique | Ranking |
|---|---|
| RFE-RBF | 3.86 |
| RFE-RF | 2.80 |
| PSO-SVM | 2.33 |
| RFE-LAP | 1.00 |

After the null hypothesis is rejected under the Quade test, a *post hoc* procedure must be used (García et al., 2010) in order to address the multiple hypothesis testing among the different models, the *p-values* should be corrected, with adjusted *p-values* (APV). There are several *post hoc* procedures under a non-parametric test to adjust *p-values* such as

Finner's test (Finner, 1993), which is easy to comprehend and offers better results than Holm, Hochberg or Hommel's, for example. Finner's procedure compares the control model (winner) with all the other models. **Table 3** presents for each technique a comparison between the *p-values*, the adjusted *p-value* with Finner's procedure and the final value achieved with Finner's procedure.

**Table 3.** Adjusting p-values with Finner's procedure

| Technique | p-value | Adjusted p-value | Finner |
|---|---|---|---|
| RFE-RBF | 0.0095 | 0.028 | 0.016 |
| RFE-RF | 0.1030 | 0.151 | 0.033 |
| PSO-SVM | 0.2270 | 0.227 | 0.050 |

Finner's procedure rejects hypothesis with value $\leq 0.033$, which means that RFE-RBF and RFE-RF are statistically significantly worse than RFE-LAP. During the cross validation process, RFE-LAP used 11 features: X5, H, S6, X4, eS6, eH, eTr3, eX5, eTr5, eTr0 and Tr0. It is interesting to notice that features Tr0, eTr0, X5, eX5, H, eH, S6, eS6 are 4 types of descriptors for both embedded and non-embedded molecular star graphs. This suggests there is a need for adding the initial sequence connectivity (in embedded graphs) in order to encode the information required for the signaling classification.

The best signaling classification model (RFE-LAP, 11 features) has been implemented as a free Web tool at http://bio-aims.udc.es/Signal-Pred.php and is available for download at http://dx.doi.org/10.6084/m9.figshare.1330133. The server is based on an HTML/PHP interface, R classification model and Python scripting and S2SNet command line version for the calculation of the topological indices.

*Prediction of proteins of unknown biological function in PDB databank*

In order to test the capability of the implemented model to predict proteins involved in signaling. A list of 3114 proteins of "unknown" function from the PDB Databank has been tested with the model. Thus, the FASTA sequences for these protein chains have been downloaded and the star graph descriptors have been calculated. These series of numbers have been introduced into our best model in order to predict the probability for these proteins to be involved in signaling. 96 PDBs with a probability over 98% to be signaling-related have been used for this analysis.

In order to check these predictions, all PDBs have been checked in a pathway database. To this end, in the first step, the Uniprot mapping Web (www.uniprot.org) has been used to obtain the Uniprot IDs for the corresponding PDB IDs (one Uniprot ID could correspond to several PDB structures). Therefore, the Uniprot IDs can be mapped for the possible signaling predicted PDBs. In the second step, the list of Uniprot IDs have been used to query the REST services of WikiPathways tool (http://www.wikipathways.org) (Kelder et al., 2009; Kelder et al., 2011; Pico et al., 2008) in order to find the pathways in which these proteins are involved. This step was performed using a Python script available from PyWikiPathways GitHub repository (https://github.com/muntisa/PyWikiPathways).

**Table 4** presents the most important human pathways linked to signaling biological activity. Important signaling pathways are presented for 3 UniprotIDs (32 PDBs) with signaling prediction greater than 98.0%. These results demonstrated the power of protein signaling prediction of the new star graph-based classification model. All the predicted signaling peptides with a sequence identity greater than 20% with signaling proteins from the positive set have been removed.

**Table 4.** Signaling predictions for human PDBs of unknown function using WikiPathways

| |
|---|
| *P00519* => PDBs: 4J9I, 4J9E, 4J9C, 4J9D, 4J9F, 4J9G, 4J9H (Signaling predictions: 99.1% – 98.7%) |
| WP2719: Fcgamma receptor (FCGR) dependent phagocytosis (Suzuki et al., 2000) |
| WP673: ErbB Signaling Pathway (McCarty, 1998) |
| WP1742: TP53 Network (Ishimoto et al., 2002) |
| WP437: EGF/EGFR Signaling Pathway (Kandasamy et al., 2010) |
| WP2516: ATM Signaling Pathway (Meuth, 2010) |
| *P00734* => PDBs: 1A46, 1A4W, 1A5G, 1A61, 1ABI, 1AD8, 1AI8, 1AWF, 1AY6, 1B5G, 1BCU, 1BMM, 1BMN, 1DIT, 1HAG, 1HXE, 1HXF, 1NRS, 1TBZ, 1TMB, 1TOM, 2HGT, 3HAT, 5GDS (Signaling predictions: 99.1% – 98.8%) |
| WP2664: Gastrin-CREB signaling pathway via PKC and MAPK (Gilon and Henquin, 2001) |
| WP2799: Regulation of Insulin-like Growth Factor (IGF) Transport and Uptake by Insulin-like Growth Factor Binding Proteins (IGFBPs) (Laursen et al., 2002) |
| *P29466* => PDB: 1IBC (Signaling predictions: 98.3%) |
| WP2763: Nucleotide-binding domain, leucine-rich repeat containing receptor (NLR) signaling pathways (Adhikari et al., 2007) |
| WP382: MAPK Signaling Pathway (Tanoue and Nishida, 2002) |
| WP2359: Parkin-Ubiquitin Proteasomal System pathway (Smith et al., 2012) |

* Use "http://www.wikipathways.org/index.php/Pathway:WP_ID" to get details about these pathways (WP_ID = WikiPathways ID)

**Conclusions**

The first classification model to predict signaling proteins has been proposed and it is based on peptide sequence information encoded into protein star graph topological indices, calculated with the S2SNet tool and Machine Learning models for feature selection.

Initially, several classification methods were tested using all the information available (composed of a total of 42 features). As result of these tests, SVM seems to be the algorithm that yields the best results. Second, the purpose was to improve the classification score using a more reduced set of features, thus two different approaches were used for feature selection: Particle Swarm Optimization with an SVM (RBF kernel as decision function) and Recursive Feature Elimination with an SVM (Laplacian and RBF kernels as decision function).

RFE-LAP selected features to improve the classification results with only 11 out of 42 features calculated with S2SNet. Therefore, these results can help predicting signaling function-related proteins using only a reduced amount of molecular information encoded into the protein sequence. Furthermore, with the new predictions, it is possible to search for new molecular targets.

The best prediction model has been tested with proteins of unknown function from the PDB Databank and the best predicted proteins have been verified with the Reactome pathway database. Several proteins have been identified as being involved in signal transduction pathways, demonstrating the prediction power of the best model. This model has been implemented as a free Web tool in the Bio-AIMS server collection.

**Acknowledgements**

This work is supported by the "Collaborative Project in Medical Informatics (CIMED)" PI13/00280 funded by the Carlos III Health Institute, from the Spanish National plan for Scientific and Technical Research and Innovation 2013-2016 and the European Regional Development Funds (FEDER). The authors would like to thank the Galician Supercomputing Centre (CESGA) for the provision of computational support and acknowledge the support from the Galician Network of Drugs R+D REGID (Xunta de Galicia R2014/025).